\documentclass[12pt,preprint]{aastex}

\newcommand{\etal}{{\em et al.}}            

\shorttitle{X-rays from Cen A ISM and Radio Lobes}
\shortauthors{Kraft \etal}

\begin{document}

\title{An Unusual Discontinuity in the X-ray Surface Brightness Profile of NGC 507: Evidence of an Abundance Gradient?}
\author{R. P. Kraft, W. R. Forman}
\affil{Harvard/Smithsonian Center for Astrophysics, 60 Garden St., MS-67, Cambridge, MA 02138}
\author{E. Churazov}
\affil{Max Planck Institute f\"{u}r Astrophysik, 85740 Garching bei M\"{u}nchen, Germany and Space Research Institute (IKI), Moscow 117810, Russia}
\author{N. Laslo, C. Jones, D.-W. Kim, M. Markevitch, S. S. Murray, A. Vikhlinin}
\affil{Harvard/Smithsonian Center for Astrophysics, 60 Garden St., MS-67, Cambridge, MA 02138}

\begin{abstract}

We present results from a 45 ks Chandra/ACIS-I
observation of the nearby ($z=0.01646$) NGC 507 group.  The X-ray surface
brightness profile of the outer region of the group is well described
by an isothermal $\beta$-profile consistent with earlier ROSAT observations.
We find a sharp edge or discontinuity in the radial surface brightness profile
55 kpc E and SE of NGC 507 covering a~$\sim$125$^\circ$ arc.
At the sharpest part of the discontinuity,
the surface brightness declines by a factor of $\sim$2
over a distance of 6.9 kpc ($\sim 15''$).
The inner and outer gas temperatures across the
discontinuity differ by only about 0.2 keV
(interior and exterior gas temperatures are 1.2 and 1.3 keV, respectively).
Spectral analysis indicates that there is a large gradient in the elemental
abundance across the discontinuity, and
comparison with the low resolution NVSS radio maps suggests that
the discontinuity is aligned with a low surface brightness radio lobe.
We conclude that the appearance of this discontinuity is the result 
of the transport of high abundance material from the center of the galaxy
by the expanding radio lobe,
similar to structures seen in {\em Chandra} observations of Perseus A and
Hydra A.

\end{abstract}


\section{Introduction}

Chandra observations of early galaxies, groups of galaxies, and clusters
of galaxies have demonstrated that the hot, X-ray emitting
coronal gas of these systems is
often not resting idly in their gravitating dark matter halos and
that many of these systems are far from relaxed.
Several different structures have been identified and studied in detail.
Chandra observations of shocks and "cold fronts" associated with sub-cluster mergers
have contributed to our knowledge of the formation of structure.
Evidence for large scale, coherent, non-hydrostatic motions of
the central regions of clusters has also been reported \citep{vik01,mar01,maz01}.
One of the most interesting, and to some degree unexpected, results
from Chandra has been the study of diverse interactions between
jets and radio lobes of radio galaxies and the ambient 
ISM/IGM/ICM \citep{fab00,fin00,mcn00,jon02,kra03}.
The study of these various motions and hydrodynamical interactions
is of fundamental importance to several astrophysical and cosmological
problems.  These observations
give us insight into the merger process and subsequent
relaxation.  By studying merger shocks, thermal histories, etc.,
we learn about the process of the formation of clusters.
In addition, understanding the radio lobe/ICM interactions may be critical in
understanding the life cycle of clusters as there may be a cyclical relationship
between cooling flows, nuclear activity, and nuclear outflow.

NGC 507 is the dominant, massive elliptical galaxy of a nearby ($z=0.01646$) group/poor
cluster (the so-called Pisces cluster).
Previous ROSAT observations have shown that NGC 507 is one of the most
X-ray luminous early type galaxies ($L_x\sim$ 10$^{43}$ ergs s$^{-1}$) in the local
universe with evidence of a large cooling flow (cooling rate of 30-40
$M_\odot$ yr$^{-1}$) at the center \citep{kim95,pao03}.
This group has an unusually large mass-to-light
ratio ($\sim$125), similar to that observed in the X-ray overluminous
elliptical galaxies (OLEGs) \citep{vik99}.
NGC 507 is also known to be an FR I radio galaxy \citep{fan86}.

In this letter, we present results from a {\em Chandra}/ACIS-I
observation of the hot ISM of this group.
We detect a sharp discontinuity in the surface brightness profile
55 kpc E and SE of the center of the galaxy spanning a 125$^\circ$ arc.
This discontinuity has been seen in XMM-Newton observations of NGC 507 as
well \citep{fab02}.  A more detailed discussion of the larger scale emission
and the central cooling flow peak based on ROSAT/HRI and {\em Chandra}/ACIS-S observations
is presented elsewhere \citep{pao03}.
Initially it appears that this discontinuity was another example of
a "cold-front" as seen in clusters of galaxies.
Such discontinuities have recently been detected in clusters of galaxies
and are commonly attributed to either infalling sub-clumps or non-hydrostatic
motions of the central core (e.g. A2142 \citep{mar00},
A3667 \citep{vik01}, A1795 \citep{mar01}).  However, there are significant problems with
this interpretation of the discontinuity observed in NGC 507.  In particular,
the temperature jump across the discontinuity is small (0.1-0.2 keV), so that the
discontinuity should not appear sharp.  In addition,
if the sharp boundary is maintained by motion of the galaxy, a large,
nearly supersonic velocity of the galaxy with respect to the larger scale
halo is required.  Spectral analysis indicates that there is
a significant jump in elemental abundance across the discontinuity.
The X-ray discontinuity is coincident with a radio lobe detected in the NVSS
survey.  We suggest that the discontinuity in the surface brightness is
due to a combination of the near-supersonic expansion of
the radio lobe and to turbulent entrainment of high abundance material from the
center of the galaxy to more distant regions caused by this expansion.

Throughout this paper, we use $H_0$=75 km s$^{-1}$ Mpc$^{-1}$, $\Omega$=0.3 and
$\Lambda$=0.7.  The observed redshift of NGC 507 then corresponds to a distance of 65.5 Mpc,
and $1'$=18.8 kpc. 

\section{Observations}

The central region of the nearby (z=0.01646, d=65.5 Mpc) NGC 507 group 
was observed twice by {\em Chandra}.  The first observation, taken Oct 11, 2000,
was made with the ACIS-S instrument, and the second, taken Jan 8, 2002,
with the ACIS-I in very faint mode.  The particle backgrounds in both
observations were examined for periods of flares,
and the average levels compared with the nominal ACIS background
rates.  We found that the first observation was contaminated
by a steady but anomalously high background, so we have used only data from the second
observation in all analyses presented in this letter.  
No flares or periods of high background were found in the
second dataset.  Very faint mode filtering was applied to this
second event file to reduce the background to the lowest possible
level.  A background event file was created from the standard ACIS 
background data appropriate for the date of the observation, and very 
faint mode event filtering applied to these data as well.
We also have obtained archival data from the XMM-Newton observation,
and have used these data to constrain spectral parameters as described below.

\begin{figure}
\plotone{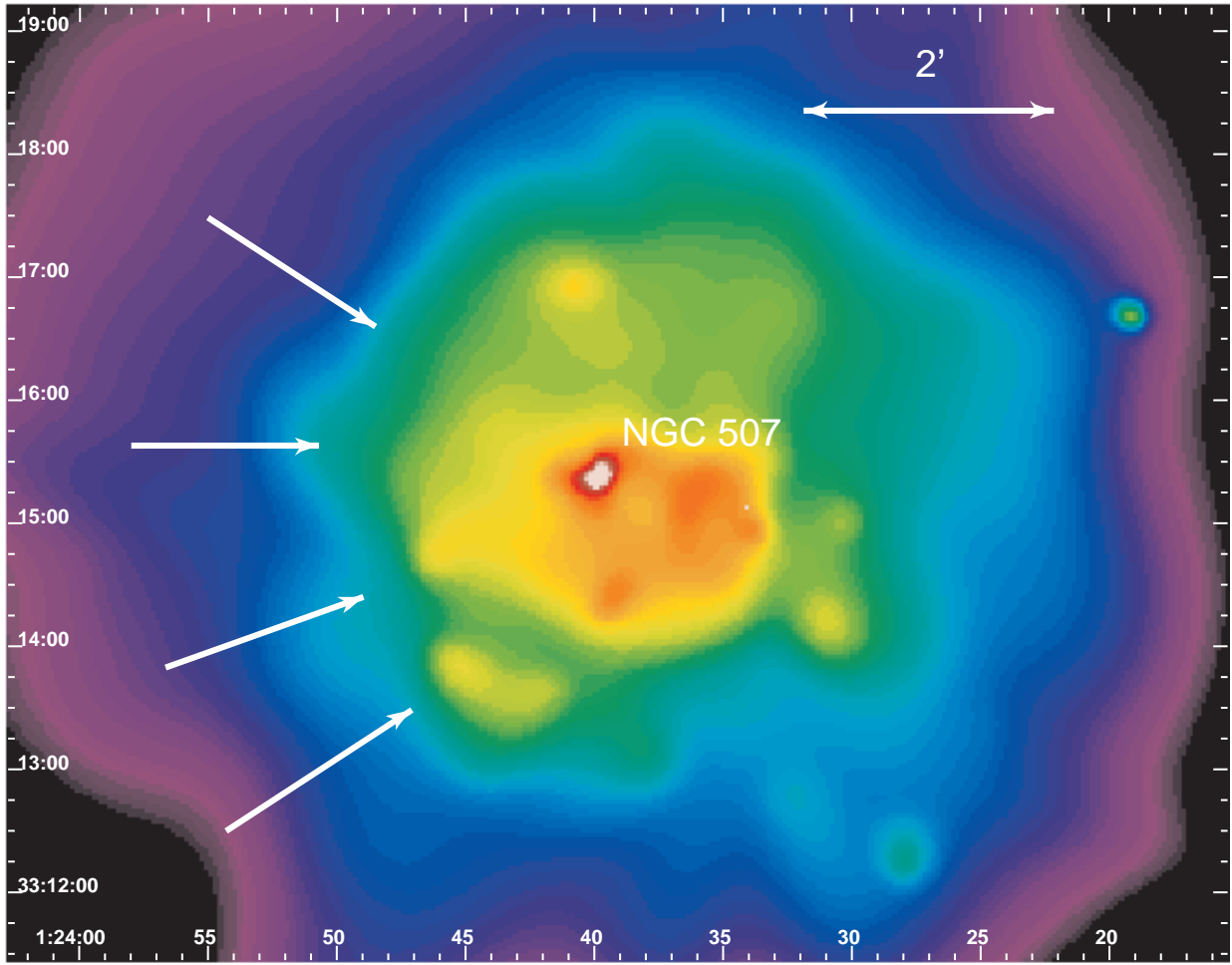}
\caption{An adaptively smoothed, exposure corrected image
of NGC 507 in the 0.5 to 2.0 keV bandpass.  The peak
of the emission is centered on NGC 507.}\label{adaptimg}
\end{figure}

A raw X-ray image in the 0.5 to 2.0 keV band is shown in Figure~\ref{rawimg}.
An adaptively smoothed, background subtracted, exposure corrected
image of NGC 507 is shown in the Figure~\ref{adaptimg}.
The complex azimuthal structure in the X-ray emission can be seen
in both figures.  The large scale coronal emission can be seen in Figure~\ref{adaptimg} with the
X-ray peak centered on NGC 507.  X-ray emission from the less massive NGC 508
is detected to the N.  Some structure in the emission is seen to the west of
NGC 507, and is most likely related to interaction between the radio lobe and
the ISM.

\section{Analysis}

\begin{figure}
\plotone{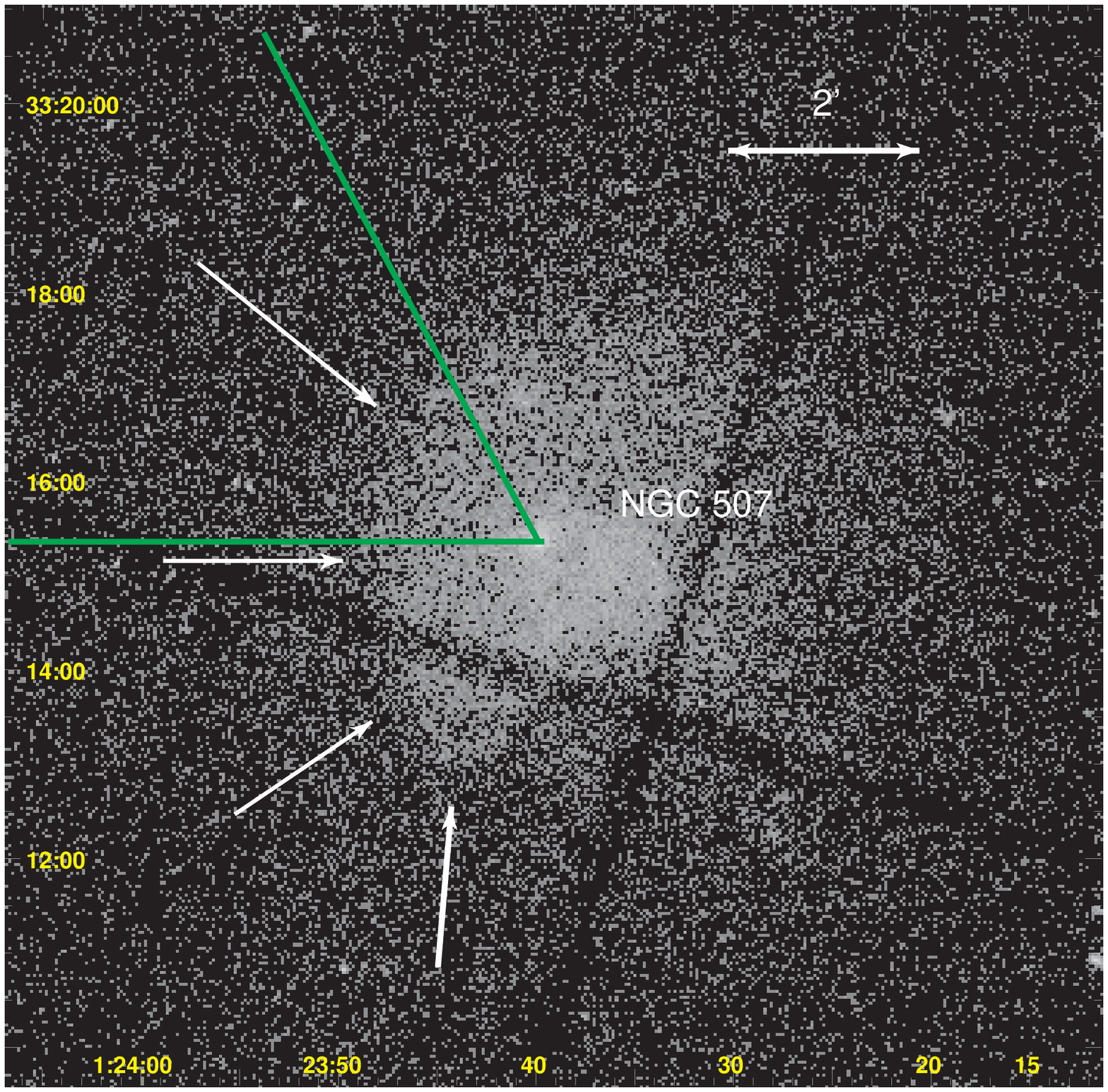}
\caption{Raw Chandra/ACIS-I X-ray image of NGC 507 in the 0.5 to 2.0
keV band.  The green wedge indicates the sector used to create
the radial surface brightness profile in Figure~\ref{sbprof}.  The white
arrows indicate the approximate position of the surface brightness
discontinuity.}\label{rawimg}
\end{figure}

\begin{figure}
\plotone{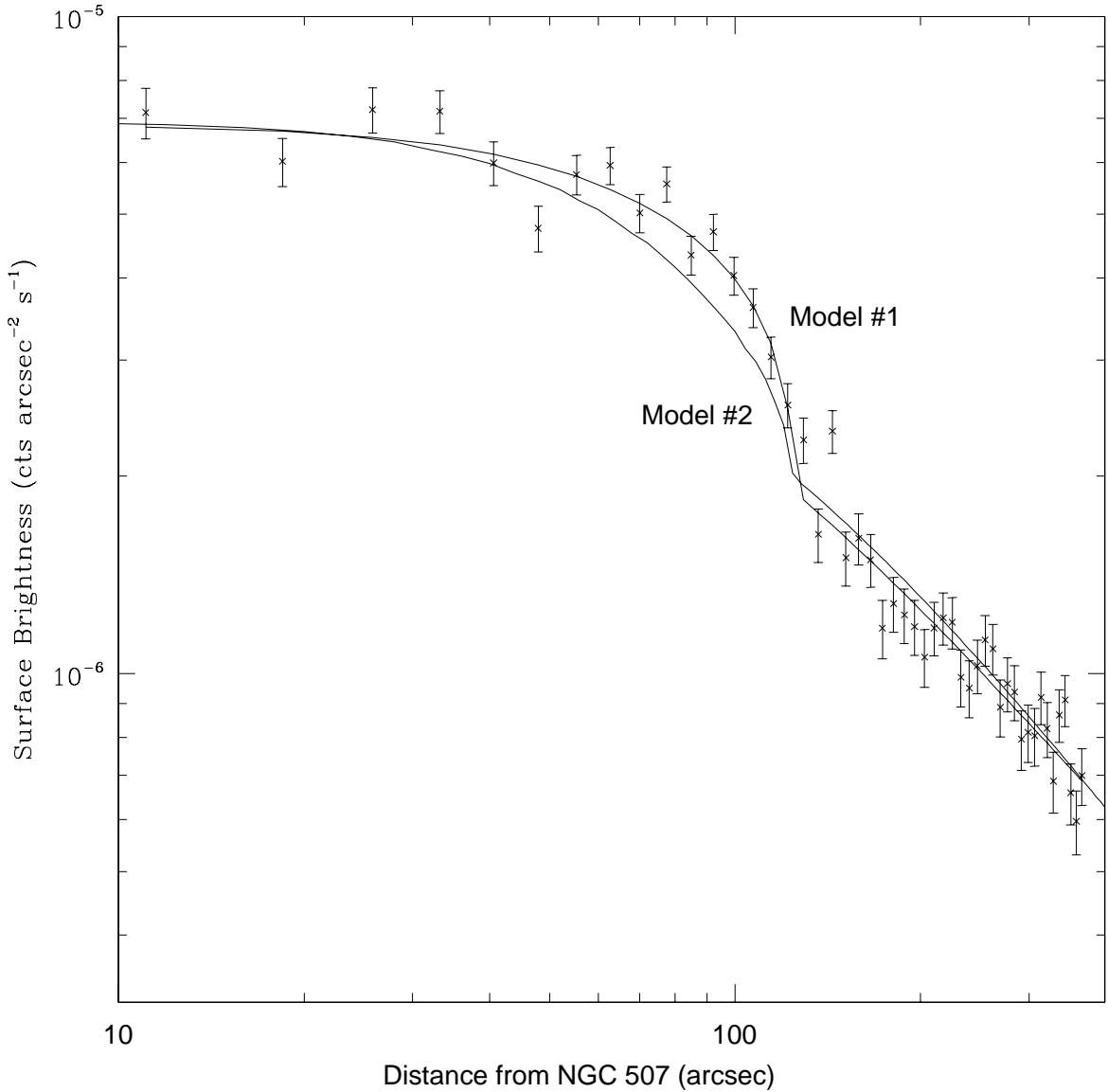}
\caption{Radial surface brightness profile of the X-ray emission in a 60$^\circ$
sector (shown in Figure~\ref{rawimg}) to the E and NE of NGC 507 in the
0.5-2.0 keV band.   The continuous curve labeled "Model \#1" is computed from a model
of the emission that consists of two components, an isothermal beta model beyond
the discontinuity and a uniform density sphere inside the discontinuity.
While reasonably consistent with the data, the model is physically unrealistic since
it attributes the discontinuity to a change in density and therefore, since the temperature
is roughly constant, to a large pressure discontinuity.
The curve labeled "Model \#2" consists of emission from a $\beta$-model
density profile but with the
emissivity of the gas interior to the discontinuity increased by a factor of 2.16
to account for the enhanced abundance and lower temperature
derived from spectral fitting (see Figure~\ref{tempfig} below).
This model can account for most of the observed structure of the discontinuity.
The expansion of the radio lobe may also be important.}\label{sbprof}
\end{figure}

One of the most striking features of Figure~\ref{rawimg}
is the sharp discontinuity in the surface brightness
approximately $120''$ to the east of the central peak spanning an angle
of approximately $\sim$125$^\circ$ from the north through the east to the south.
The radial surface brightness profile in the 0.5 to 2.0 keV
band of the 60$^\circ$ wedge shown in Figure~\ref{rawimg}
is plotted in Figure~\ref{sbprof}.
The discontinuity appears to be sharpest to the SE, but in creating
the surface brightness profile in Figure~\ref{sbprof}, we chose
the wedge to the E and ENE to avoid uncertainties associated with the CCD chip gap
toward the SE and NGC 508 to the N.

The temperature of the gas around NGC 507 was determined by two independent methods.
First, we divided a 120$^\circ$ wedge (twice the angle of the sector
shown in Figure~\ref{rawimg} to improve statistics) into eight regions,
three to the west of the discontinuity, one straddling the
discontinuity, and four to the east.  A single temperature
absorbed APEC model was fit to both the {\em Chandra} and
the XMM-Newton data (independently) with the abundance as a free parameter.
The absorption was held fixed at the galactic value ($N_H$=5.5$\times$10$^{20}$ cm$^{-2}$).
The lower limit of the spectral fits was restricted to 0.8 keV to avoid uncertainties
in the ACIS-I response due the build up of contaminants, although a
correction also was applied to the response to account for this effect.

The best fit Chandra and XMM-Newton/MOS temperatures and abundances
with 90\% confidence intervals as a function of distance
from the center of NGC 507 are plotted in Figure~\ref{tempfig}.
Figure~\ref{tempfig} shows only a small
(0.1-0.2 keV) difference in the temperature between the region
interior and exterior of the discontinuity.
The abundance increase from 0.3$\pm$0.1 to 0.6$\pm$0.1
towards the center is statistically
significant at the 90\% confidence level.
If the abundance is held constant, the temperature difference between
the interior regions and the exterior regions increases by
a only small ($\sim$0.15 keV) amount.

\begin{figure}
\plotone{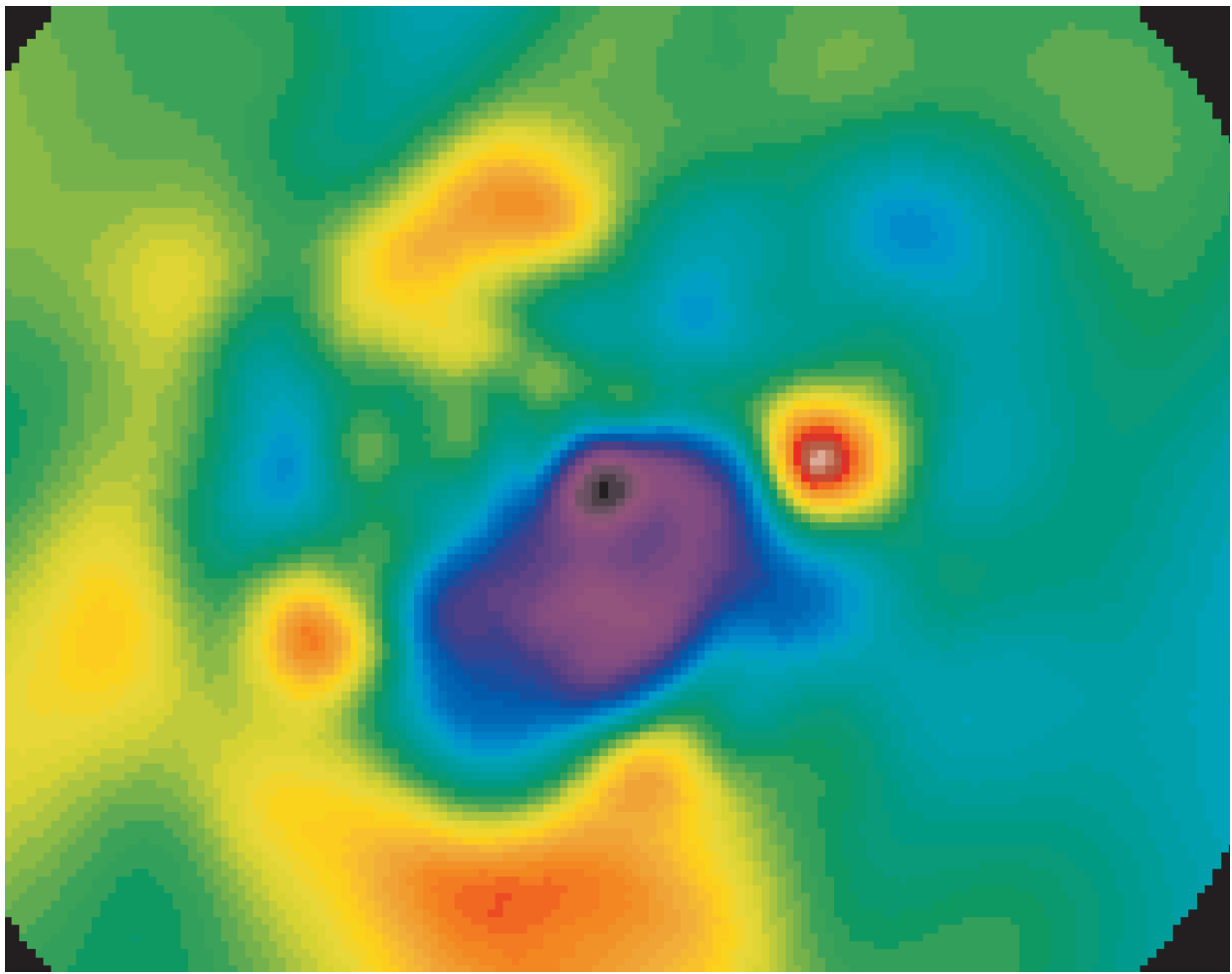}
\caption{XMM-Newton/MOS temperature map of NGC 507.  The difference in temperature
between the blue/green and the red/yellow is approximately 0.2 keV.}\label{tmapfig}
\end{figure}

We also created a temperature maps from both data sets
using the algorithm described in \citet{chu96}.  
The map created from the XMM-Newton/MOS data
is shown in Figure~\ref{tmapfig} and
supports the conclusion of the spectral fitting
that there is little change in temperature just across
the discontinuity, but a small temperature decrease toward the center. 
Interestingly the temperature is lowest in the
region to the S/SE where the discontinuity appears sharpest.

\section{Interpretation}

If the abundance is roughly constant across the discontinuity,
there must be a large pressure jump due to the large density discontinuity because
the observed temperature jump across the boundary is small.
We consider three possible interpretations to explain the observed
surface brightness feature.  In the first
case, we consider the possibility that the edge is caused by motion
of NGC 507 relative to the larger scale group dark matter potential.
Second, we consider a scenario where the pressure across the boundary
is balanced by an unseen relic radio lobe left over from an
earlier (and more powerful) epoch of nuclear activity.  Finally, we investigate
the possibility that the edge is due to the subsonic expansion of
the relatively weak lobe currently observed lying interior to the
discontinuity.  This expansion has pushed higher elemental abundance gas
out from the central region.  The appearance of the discontinuity is
then a consequence both of the compression of the gas due to
expansion of the lobe and to the increase in the abundance across the
discontinuity.

One interpretation commonly invoked
to explain the sharp discontinuities observed in 
galaxy clusters is that the central object is moving with a significant velocity
relative to the larger halo \citep{vik01,mar01}.  These are the so-called
cluster "cold fronts".  There are several problems with this scenario
for NGC 507.  First, the implied velocity must be very large.
If the pressure difference between the interior and exterior regions is balanced
by ram pressure, the implied velocity is approximately 450 km s$^{-1}$
(Mach 0.7) relative to the ambient medium.  Such a large velocity
relative to the dark matter halo is not reasonable since
NGC 507 is the largest galaxy in the group and lies at
or near the center defined by the outer X-ray isophotes.
The large scale gravitating mass
of the NGC 507 group can be estimated from the parameters of the $\beta$-profile
assuming the larger scale gas is in hydrostatic equilibrium.
If we consider NGC 507 to be an undamped harmonic oscillator in this dark matter potential
moving with a velocity of 450 km s$^{-1}$,
it should reach a distance of $\sim$75 kpc from the center before falling
back.  Since there are no other known instances of the largest galaxy in a group
or cluster being so far from the center of the gravitational
potential, we consider this possibility to be unlikely.

A second significant complication with this interpretation arises from
hydrodynamic considerations.  The pressure gradient
is due largely to a density difference, not a temperature difference.
There is no evidence for supersonic motions (i.e. there is no
hot, shock heated shell in front of the gas which would
be moving supersonically).  This implies that the motion is subsonic (consistent with the above
analysis), but since the pressure difference is due only to differences in
density, the boundary should not appear sharp, as there must be a
continuous and gradual pressure and density transition between the two regions.
If the pressure difference were due to a temperature discontinuity (such
as in A3667 \citep{vik01}), a sharp boundary could be observable because
the surface brightness is proportional to $n^2$.
It is conceivable that a much smaller region of the
galaxy is actually moving, and the discontinuity is material that is piled
up in front of the stagnation point.
A large velocity would still be required to contain this material.

\begin{figure}
\plotone{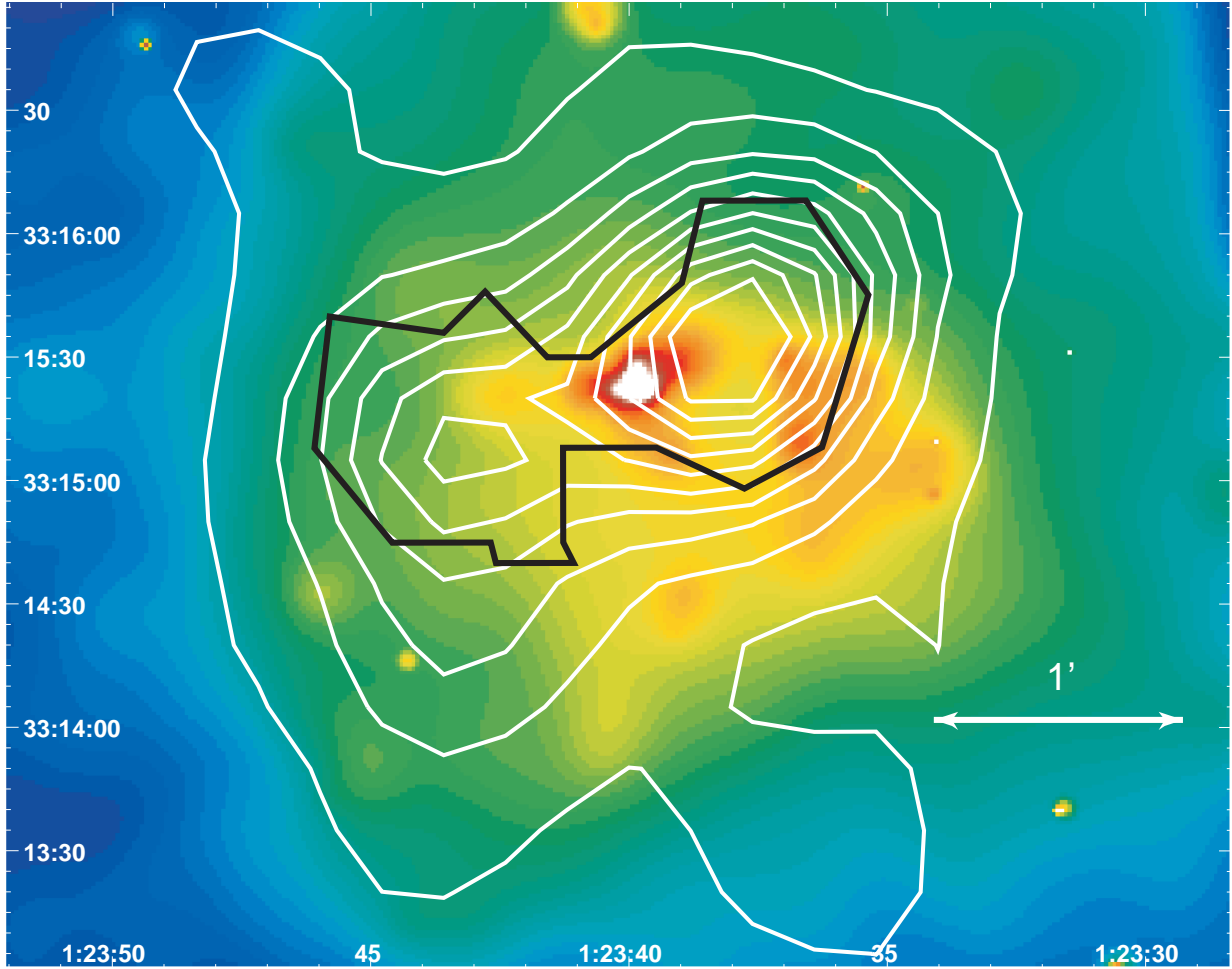}
\caption{NVSS (white - 1.4 GHz - Condon \etal~1998) and earlier, higher
resolution VLA (black - 1.4 GHz - de Ruiter \etal~1986) radio contours overlaid onto an
adaptively smoothed Chandra image.  Note the close correspondence between
the boundaries of the NVSS radio emission and the X-ray surface brightness
discontinuity to the S/SE of the nucleus.  The spatial resolution ($45''$ FWHM restoring beam) and
sensitivity (0.45 mJy at 1.4 GHz) of the NVSS data is rather poor, but this
data suggests that there is, in fact, a relationship between the lobe
and the discontinuity.  The higher resolution map ($\sim 15''$ beam FWHM) indicates
a radio lobe inside the discontinuity.}\label{nvss}
\end{figure}

For the sharp boundary to exist in the absence of motion,
additional pressure support could exist beyond the discontinuity.
An unseen relic radio lobe could provide the needed pressure support beyond the
discontinuity.  NGC 507 is a known radio source and has been
classified as an FR I radio galaxy (Fanti \etal~1986).  Data from
two radio observations are shown in Figure~\ref{nvss} overlaid
onto an adaptively smoothed Chandra image.
These radio observations detected a double lobe structure lying along the
east-west axis of the galaxy.  The eastern (weaker) radio lobe is contained
inside the discontinuity.  The western lobe is contained within some of
the complex emission to the west and may be responsible for
the complex morphology \citep{for01}.  No evidence for radio emission beyond the X-ray
discontinuity has been reported.

Is it feasible that such an unobserved, relic lobe is providing pressure
support beyond the observed discontinuity?  
Relativistic electrons with $\gamma\sim$5$\times$10$^2$ in a magnetic field of 50 $\mu$G
(a typical value for the lobes of FR II radio galaxies)
would radiate at $\sim$10 MHz, which is unobservable.  The density of
such electrons can be estimated by requiring the pressure of
the lobe to balance the pressure difference across the discontinuity.
Assuming that the lobe has a $\sim$ 50 kpc radius to cover the observed
discontinuity, the total energy
in the lobe would be about 5$\times$10$^{60}$ ergs.  This number is
a factor of a few larger than the total energy in one of the lobes
of the FR II radio galaxy Cyg A \citep{car91}, and would imply an energy deposition rate
of approximately 10$^{46}$ ergs s$^{-1}$ for 10$^7$ years.
The luminosity of IC scattered CMB photons would be considerable
($\sim$10$^{9}$ L$_\odot$), but peaking in the EUV ($\sim$30 eV),
again difficult to observe.
The existence of such an energetic but otherwise unobservable radio
relic is not ruled out by this simple energy argument, but is
highly unlikely.

As a final possibility, we consider a scenario where the inflation/expansion of
the known radio lobe is creating the discontinuity.  Such a scenario can
explain all the observed features, and has recently
been investigated theoretically by \citet{rey00} to understand
{\em Chandra} observations of cavities and cool shells seen around the lobes
of several other FR I radio galaxies.  We suggest that there is a low
surface brightness radio lobe interior to the discontinuity, expanding to the S/SE.
Such a lobe is hinted at in the NVSS radio map (see Figure~\ref{nvss}).
The relationship between the discontinuity and the smaller scale
radio lobe presented in \citet{fan86} is not clear.
In this scenario, the subsonic expansion of the radio lobe, whether
due to ongoing nuclear activity or by buoyancy if the central
source has shut off, gently pushes
the lower temperature, higher abundance material from the cooling center
of the galaxy out to more distant regions \citep{chu01,nul02}.

\begin{figure}
\plotone{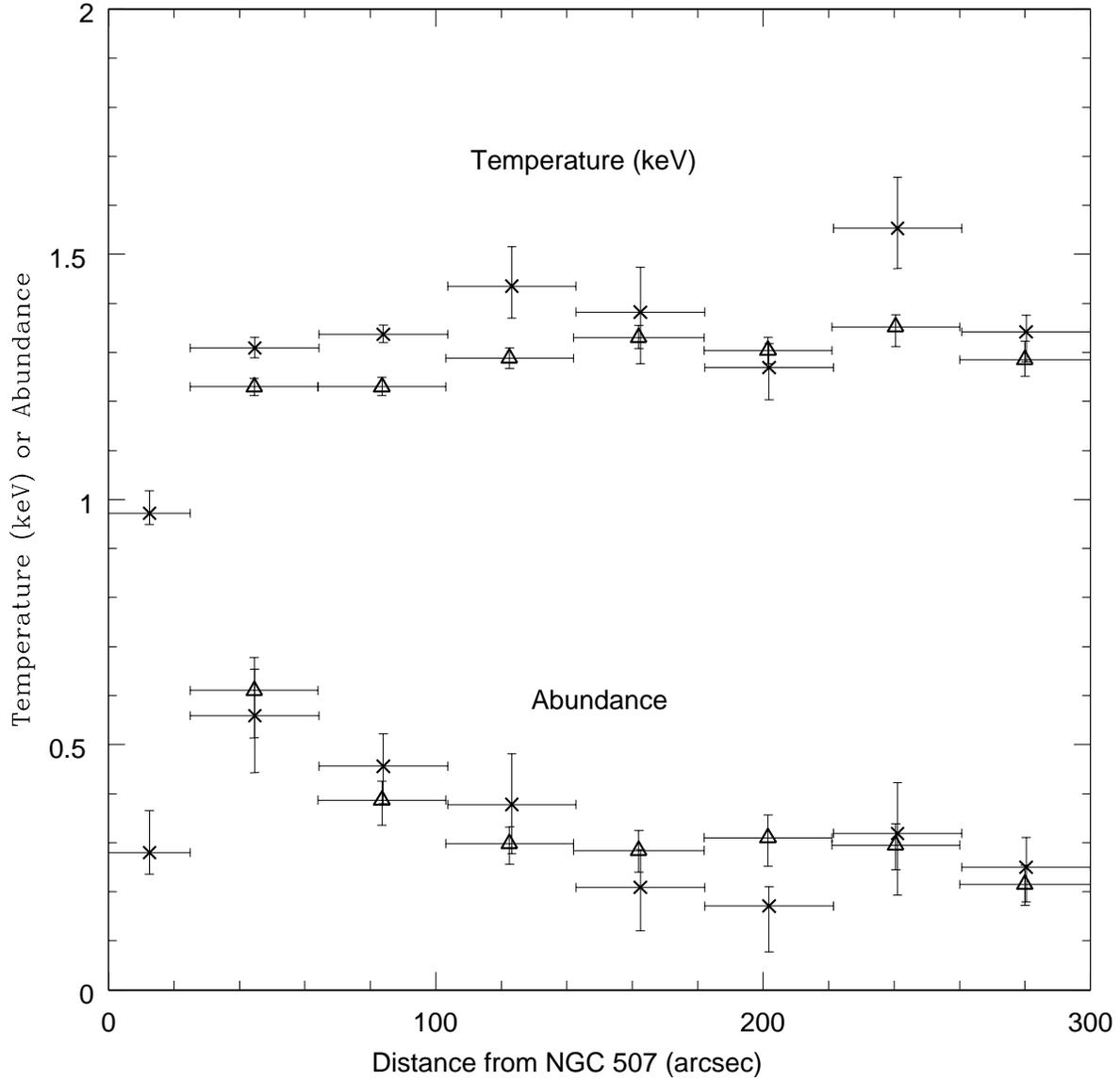}
\caption{The best-fit temperature and elemental abundances in a 120$^\circ$ sector
to the SE, E, and NE of NGC 507.  The Xs and the triangles correspond to the Chandra/ACIS-I
and XMM-Newton/MOS fits, respectively.  The top set of points is the temperature (keV), and the
bottom set is the elemental abundance (fractional relative to solar).  All error bars
are 90\% confidence intervals for one interesting parameter.}\label{tempfig}
\end{figure}

We hypothesize that the sharp appearance of the discontinuity is due
to the combination of the motion of the radio lobe and
to a significant abundance discontinuity across the interface.
At the observed $\sim$ 1.2-1.3 keV temperature of NGC 507,
the emissivity of the gas is a strong function of
the heavy element abundance, and there is an abundance gradient
from the halo to the core (see Figure~\ref{tempfig}).
Material entrained from the central regions of
NGC 507 would have higher abundance than that in the
halo and would naturally give a sharp boundary if the emission were due
to a cap surrounding a radio lobe.
That is, the sharp abundance discontinuity (shown in
Figure~\ref{tempfig}) would naturally produce a discontinuity in the surface
brightness, and at the same time maintain pressure equilibrium without the
need for a large temperature jump.

To demonstrate the effects of the enhanced elemental abundance
interior to the discontinuity, we consider a model for the surface brightness where
the emissivity of the gas is determined by
the extreme values of temperature and abundance in
Figure~\ref{tempfig}.
That is, the gas interior to the discontinuity has an abundance more than twice that
of the material beyond the discontinuity, but with a 20\% lower temperature.
We assume that the gas density is described by a single $\beta$-profile,
and that there is no pressure discontinuity across the
surface brightness discontinuity.
The emissivity of the interior material is enhanced by a factor of 2.16
due to its higher abundance and lower temperature.
We have overplotted the surface brightness profile from this
model onto Figure~\ref{sbprof}.

Such an idealized model closely approximates the observed surface brightness
profile and can account for most, but not all, of the
emission interior to the discontinuity.  
The remaining difference in the surface brightness (between
the data and the model) can be attributed to compression of
the gas by the inflation of the radio lobe.  
We have increased the density, and therefore pressure, of the interior region to match the
surface brightness profile, and estimate that
an expansion velocity of the lobe on the order of 100 km s$^{-1}$ is
required to match the profile in detail and balance the increased pressure.
We therefore conclude that a model combining the
effect of the enhanced abundances and compression from subsonic
motion can easily match the observed surface brightness profile.
Further detailed radio observations are required to constrain the relationship
between the lobe and this discontinuity.

{\em Chandra} has now detected more than a dozen examples of active
galaxies "blowing bubbles" into the ICM (\citet{bla01, mcn00} among
many others).  In most cases, there is no evidence of shocks being
driven into the IGM/ICM.  This is clearly the case with NGC 507
as there is no evidence of hotter gas associated with either the
discontinuity or any of the radio components.
The inflation of the lobe has however disrupted the gas in the central
region and may provide an efficient mechanism to mix the higher abundance material at the center
with the lower abundance material (presumably much older) in the halo.

\section{Conclusions}

We have observed a sharp discontinuity in the X-ray surface brightness profile in
a Chandra observation of NGC 507.
We conclude that this discontinuity is most likely not a "cold front" as has been
observed in many clusters, and this is not due to non-hydrostatic motion of the group core.
It is also highly improbable that this discontinuity is supported by
an unseen, relic radio lobe from an earlier epoch of galaxy activity.
The appearance of this discontinuity is probably due to two factors.  First,
the subsonic expansion/inflation the radio lobe is compressing the ambient ISM.
Second, there is an abundance gradient across the boundary that creates
a discontinuity in the surface brightness.  We suggest that the inflation of
the radio lobe has pushed up and/or entrained higher abundance material from the
central regions within NGC 507 into the group halo.
It is unclear whether this is a common phenomenon or not because it is
most readily observable only in relatively cool ($k_B T <$ 2 keV) galaxies and clusters
where the emissivity of the gas is a strong function of the abundance.
This type of disruption/mixing of the central, high
abundance region of galaxy groups and clusters by jets/lobes may be common, but would be 
difficult to detect in high temperature systems where abundance variations
would not produce changes in surface brightness but instead would require detailed
abundance maps for their detection.
There is at least one other object, A576, that shows evidence for
a surface brightness discontinuity that is at least partially
attributable to a gradient in the elemental abundance (Kempner and David, in preparation).
If this is indeed a common phenomenon, the infall velocity of cool subclusters or
the expansion velocity of radio lobe/ICM interactions would be overestimated, perhaps
by as much as a factor of a few.

\acknowledgements

This work was supported by NASA contracts NAS8-38248, NAS8-39073, the
Chandra X-ray Center, and the Smithsonian Institution.

\clearpage

\clearpage

\clearpage

\clearpage

\clearpage

\clearpage

\end{document}